\newcommand{\be}{\begin{equation}}
\newcommand{\ee}{\end{equation}}
\newcommand{\bea}{\begin{eqnarray}}
\newcommand{\eea}{\end{eqnarray}}
\newcommand{\ba}{\begin{array}}
\newcommand{\ea}{\end{array}}

\def\bbox{{\,\lower0.9pt\vbox{\hrule \hbox{\vrule height 0.2 cm
\hskip 0.2 cm \vrule height 0.2 cm}\hrule}\,}}
\newcommand{\dsl}{\pa \kern-0.5em /}

\def\psizero{\psi^{(0)}}

\documentclass[aps,twocolumn,showpacs]{revtex4}
\usepackage{graphicx}

\def\gammax{ \gamma ^{\hat 1 } }
\def\gammat{ \gamma ^{\hat 0} }
\def\gammatx{ \gamma ^{\hat 0 \hat 1} }

\def\psidagger { \psi ^{\dagger} }

\def\mass {{\cal M}}

\begin{document}

%\draft

%<<<<<<<<<<<<< TITLE >>>>>>>>>>>>>>>%
%\title{On the Positivity of Y-ADM Mass per Unit length} 
\title{Positivity Bounds for the Y-ADM Mass Density}

%<<<<<<<<<<<<< AUTHOR >>>>>>>>>>>>>>>%
\author{David Kastor$^{(1)}$, Tetsuya Shiromizu$^{(2,3,4)}$,  
Shinya Tomizawa$^{(2)}$ and
Jennie Traschen$^{(1)}$ }

%<<<<<<<<<<<<< ADDRESS >>>>>>>>>>>>>>>%

\affiliation{$^{(1)}$Department of Physics, University of Massachusetts, Amherst, MA 01003, US}

\affiliation{$^{(2)}$Department of Physics, Tokyo Institute of Technology, 
Tokyo 152-8551, Japan}

\affiliation{$^{(3)}$Department of Physics, The University of Tokyo,  Tokyo
113-0033, Japan}

\affiliation{$^{(4)}$Advanced Research Institute for Science and Engineering, 
Waseda University, Tokyo 169-8555, Japan}

%<<<<<<<<<<<<< DATE >>>>>>>>>>>>>>>%
\date{\today}

%======================================%
%<<<<<<<<<<<<< ABSTRACT >>>>>>>>>>>>>>>%
%======================================%
\begin{abstract}
Killing-Yano tensors are natural generalizations of Killing vectors to arbitrary rank anti-symmetric
tensor fields.   It was recently shown that Killing-Yano tensors lead to conserved gravitational charges,
called Y-ADM charges.  These new charges are interesting because they measure {\it e.g.} the mass density of a $p$-brane, rather than the total ADM mass which may be infinite.
In this paper, we show that the spinorial techniques used by Witten, in his proof of the positive energy theorem, may be straightforwardly extended to study the positivity properties of the Y-ADM mass density for $p$-brane spacetimes.   Although the resulting formalism is quite similar to the ADM case, we show that establishing a positivity bound in the higher rank Y-ADM case requires imposing a condition on the Weyl tensor in addition to an energy condition.  We find appropriate energy conditions for spacetimes that are conformally flat or algebraically special, and for spacetimes that have an exact Killing vector along the brane. Finally we discuss our expression for the Y-ADM mass density from the Hamiltonian point of view. 
\end{abstract}


\maketitle
%\vskip1cm

%======================================%
%<<<<<<<<<<<< SECTION I  >>>>>>>>>>>>>>%
%======================================%
%\baselineskip25pt
%\label{sec:introduction}
%\section{Introduction}

\section{introduction}

Conserved gravitational charges are associated with the asymptotic symmetries of a spacetime.
For example, the ADM mass and angular momentum correspond to asymptotic time translation and rotation Killing vectors respectively. In $D$ spacetime dimensions these charges are
given by integrals over a $(D-2)$-sphere at  spatial infinity. 
The electric charge carried by matter and black holes in the spacetime interior is, of course, also given
by an integral over a $(D-2)$-sphere.   Charge conservation allows both these integrals to be 
evaluated at an arbitrary time.

For $p$-branes this coincidence no longer holds.  
The ADM mass, in this case, is given by an integral over a cylinder $R^p\times S^{D-(p+2)}$
at transverse spatial infinity that encloses the entire brane -  
{\it i.e.} the integral includes directions parallel to the brane.  A $p$-brane world-volume naturally couples to a  $(p+1)$-form gauge potential.  The electric charge of the brane is
given by an integral over an $S^{D-(p+2)}$ that encloses only a single point on the brane world-volume - {\it i.e.} the integral excludes directions parallel to the brane.   Charge conservation allows the ADM integral to be evaluated at
an arbitrary time, while the electric charge integral may be evaluated at an arbitrary position along the brane in \textit{space or time}.

Symmetry may be restored to the formulation of  gauge and gravitational 
 charges for $p$-branes  by  the introduction of new gravitational charges,
called Y-ADM charges,  associated with Killing-Yano tensors \cite{KT}.
A Killing-Yano tensor is an antisymmetric tensor field of arbitrary rank  that satisfies a natural generalization of Killing's equation \cite{Yano}.   A rank $1$ Killing-Yano tensor is simply a Killing
vector.  It was shown in reference \cite{KT} that the Abbott-Deser (AD) construction  of the ADM charge associated with an asymptotic Killing vector \cite{AD} may be generalized to give a conserved charge 
associated with an asymptotic Killing-Yano tensor of higher rank.  The resulting Y-ADM charge is
given by an integral over an $S^{D-(p+2)}$ at transverse spatial infinity.  Like the electric charge
carried by a brane, it is  independent of time and also independent of translations of the surface of integration  parallel to the brane.  

For a $p$-brane spacetime, we focus on the rank $(p+1)$ asymptotic Killing-Yano tensor given by the anti-symmetric product of translations in the directions parallel to the 
world-volume of the brane.  For a particle this is just the asymptotic time translation Killing vector 
and the corresponding charge is the ADM mass.  For a general $p$-brane, we call 
the corresponding charge the \textit{Y-ADM mass density}, which we will denote by $\mass$ below.

One very important result associated with the ADM mass is the positive energy theorem  \cite{Schon:1979rg}.  It is natural to ask if any sort of positivity result holds for the Y-ADM mass density $\mass$?
In this paper we investigate this question, making use of the spinor techniques of Witten \cite{Witten} and Nester \cite{Nester}.   We focus on the  rank $2$  case corresponding to a $1$-brane, or string.  

Our central result is a spinorial boundary integral expression for the Y-ADM mass density 
$\mass$.  Assuming that the spatial slices have no interior boundaries, {\it e.g.} at horizons, Stokes theorem relates this boundary integral to 
a volume integral, whose integrand is the sum of three terms.  The first of 
these terms involves the stress-energy tensor; the second term is a positive definite
expression quadratic in 
the derivative of the spinor field; the third term involves certain components of the Riemann tensor.
The first two terms are similar to those that appear in Witten's proof of the positive energy theorem 
\cite{Witten}, in which an energy condition is sufficient to ensure positivity of the mass.
The third term, however, cannot be written directly in terms of the stress-energy tensor and - hence -
an energy condition is not sufficient to imply positivity of $\mass$.  
We discuss two types of further conditions under which positivity nevertheless holds: 
(i) If the spacetime is conformally flat or algebraically special and (ii) if the spacetime has an exact translational symmetry along the brane.

The paper is organised as follows.  Section II presents various preliminaries - 
definitions and results relating to Killling-Yano 
tensors, transverse asymptotically flat spacetimes, and spinors. 
In Section III, we briefly review the construction of Y-ADM charges \cite{KT}
using the techniques of Abbott and Deser \cite{AD}. 
We show, in particular for the rank $2$ case, that 
the Y-ADM mass density $\mass$ is intrinsic in the sense that it depends only on
quantities intrinsic to the codimension $2$ slice transverse to the string - a result that has a 
close parallel in the rank $1$ case, in the formula  for the ADM mass.
Our spinorial construction for Y-ADM charges is presented in Section IV.
In Section V, we demonstrate positivity of the mass density $\mass$ under the special 
conditions mentioned above.  
Section VI is a discussion of 
preliminary results on the Y-ADM mass density as a $p$-brane Hamiltonian and some
concluding remarks.

\section{Preliminaries}

\subsection{Killing-Yano Tensors}

A Killing-Yano tensor \cite{Yano} is a rank $n$ 
rank antisymmetric tensor field $f^{a_1\cdots a_n}=f^{[a_1\cdots a_n]}$ 
satisfying 
%===========<Equation>===========%
%
\begin{eqnarray}
\nabla_{(a_1}f_{a_2)a_3\cdots a_{n+1}}=0.
\end{eqnarray} 
%
%================================%
For rank $n=1$ this condition reduces to Killing's equation.  
Flat spacetime has the maximal number of Killing-Yano tensors of each rank.  
In this paper we will focus on the case of rank $n=2$.  For flat 
spacetime in Cartesian coordinates, a basis for rank $2$ Killing-Yano tensors is 
given by the translational tensors
\be\label{trans}
%f^{(ab)}
f=  dx^a\wedge dx^b,
\ee
together with the rotational tensors
\be
%f^{(abc)}
f=  x^a dx^b\wedge dx^c + x^b dx^c\wedge dx^a + x^c dx^a\wedge dx^b. 
\ee

\subsection{Transverse asymptotically flat spacetimes}\label{transversesection}

The notion of transverse asymptotic flatness in $D$-dimensions, as discussed in 
reference \cite{Townsend:2001rg},  is motivated by considering $p$-brane spacetimes.  
Such spacetimes become flat only as we approach  transverse spatial infinity
 in the $D-(p+1)$ directions transverse to the brane; not by moving along the $p$ spatial 
 directions parallel to the brane \footnote{In this paper we take the $p$ spatial directions tangent to the brane will be taken to be either infinite in extent or periodically identified on a torus.  It
 might also be interesting to consider wrapping the tangent directions on nontrivial cycles of more 
 general reduced holonomy spaces. }.  

Write the full spacetime metric as 
$g_{ab}=\eta_{ab} + h_{ab}$,
where $\eta_{ab}$ is the flat Minkowski metric.  Let $(x^\mu,x^I)$ with $\mu=0,1,\dots,p$ and $I=p+1,\dots, D-1$ be Cartesian coordinates in the asymptotic region and let $r^2=\delta_{IJ}x^I x^J$.  Transverse asymptotic flatness requires that in the limit $r\rightarrow\infty$ the components of 
$h_{ab}$ fall-off as
\be\label{falloffrates}
h_{\mu\nu}, h_{IJ} \sim {\cal O}({1\over r^{D-(p+3)}}),  \quad 
h_{\mu I}\sim {\cal O}({1\over r^{D-(p+2)}}).
\ee
Note, in particular, that  cross terms  between the directions tangent and transverse to the brane
are required to fall off one power faster than the other components.  

In this paper we will focus on the case of $1$-branes, or strings.   In addition to transverse
asymptotically flat boundary conditions with $p=1$, we will 
assume that the spacetime interior  can be foliated
by codimension two spatial submanifolds, which we will think of as transverse to the brane. 
At infinity these spatial submanifolds are taken to be aligned with the 
$(x^2,\dots,x^{D-1})$ hyperplane.  
The spacetime metric may then be written as  
 $g_{ab}=L_{ab} + q_{ab} $, with $L_a^{\ b} q_{bc}=0$,
where $q_{ab}$ is a Euclidean metric on the $(D-2)$ dimensional  submanifolds
and $L_a^{\ b}$ projects onto the directions normal to these submanifolds. 
It will also prove useful to be able to further split the metric as
\be\label{metric}
g_{ab}= -n_a n_b + y_a y_b +q_{ab}
\ee
where $n^a$ and $y^a$ are normal vectors to the submanifolds satisfying
$n_a n^a =-1 ,\   y_a y^a =1 ,\   n_a y^a =0$.  At infinity,  we take 
\be
n^a=(\partial/\partial x^0)^a,\qquad y^a= (\partial/\partial x^1)^a.
\ee
The asymptotic Killing-Yano tensor used to define the Y-ADM mass density of a string in the $x^1$ direction has non-zero coomponents $f^{01}=-f^{10}=1$.  At infinity, we can then also write
\begin{equation}\label{stringyano}
f^{ab} = n^a y^b -y^a n^b.
\end{equation} 

The construction of the Y-ADM mass density for a string takes place on a codimension two volume
$V$  with the two normals $n^a$ and  $y^a$, and  metric $q_{ab}$. The
embedding of $V$ in the full spacetime is described by the extrinsic curvature
tensor $K_{ab}^{\ \ c } =q_a ^{\ m} q_b ^{\ n} \nabla _m q_n ^{\ e}$. 
The indices $a,b$ are tangent
to $V$, and the index $c$ is normal, {\it i.e.} $ q_{cd}  K_{ab}^{\ \ c} =0$.  
The extrinsic curvature $K_{ab}^{\ \ c} $  is symmetric on the indices $a,b$, a result
that follows from Frobenius' Theorem.  The Gauss-Codazzi relations, which will
be used in subsequent calculations, are
\be\label{gcbig}
 q^{\ e} _a q^{\ f} _b q^{\ r} _c q^d _{\ s} R_{efr}^{\ \ \ s} [g] =  R_{abc}^{\ \ \ d} [q]
 - K_{ca }^{\ \ e} K_{b\  e}^{\ d} +   K_{cb }^{\ \ e}  K_{a\  e}^{\ d} 
 \ee
The extrinsic curvature can be decomposed as $K_{ab}^c =\alpha _{ab} n^c +\beta _{ab}x^c
$,
where $\alpha _{ab} = q_a^c q_b^d \nabla _c n_d$, and  
$\beta _{ab} = - q_a^c q_b^d \nabla _c y_d$.

\subsection{Spinors and Killing-Yano tensors}

We are interested in extending Witten's spinorial techniques to study the positivity of Y-ADM charges.
Hence, we assume the existence of a spinor field $\psi$ in the spacetime.   At transverse spatial infinity
we will assume that $\psi$ approaches one of the constant spinors $\psi ^{(0)}$ of flat spacetime.
We will make use of the  quantity $f_{ab}= -\psidagger \gammatx \gamma _{ab} \psi$ below, where 
$\gammat \equiv -n_a \gamma ^a$ and $\gammax \equiv x_a \gamma ^a$.  In the 
asymptotic limit, where $\psi$ is replaced by $\psi ^{(0)}$, the tensor $f_{ab}$ is a Killing-Yano tensor.  Note that  we can only construct the translational Killing-Yano tensors (\ref{trans}) of the background flat spacetime in this way.  The particular Killing-Yano tensor of interest to us (\ref{stringyano}) is of this form.

%Following Witten's proof of the positive mass theorem using spinors, we assume 
%the existence
%of a spinor field $\psi$ in the spacetime. The boundary condition on $\psi$ is 
%that it approach
%a covariantly constant spinor $\psi ^{(0)}$ at infinity, $\nabla ^{(0)} _a \psi ^{(0)} =0$. 
%This is a consistent
%boundary condition since  asymptotically the   spacetime
%is flat, and has constant spinors. Then  $f_{ab}= -\psidagger \gammatx \gamma _{ab} \psi$.
% is an asymptotic Yano tensor, since  asymptotically $\nabla ^{(0)} _c f _{ab} =0$. 
%Here $\gammat \equiv -n_a \gamma ^a , \gammax \equiv x_a \gamma ^a$.
%Note that using spinor bilinears, we can
%only form twist free Yano tensors, $i.e.$, ones which satisfy $\nabla _{[c} f_{a]b}=0$. Put differently, the two form is harmonic as well as Yano.

\section{Y-ADM Charges}\label{YADM}

We begin by briefly reviewing the construction of Y-ADM charges in transverse 
asymptotically flat spacetimes \cite{KT}.  We will continue to focus on the case of rank $2$ Killing-Yano tensors.  The
general case is considered in \cite{KT}.
The construction is an 
extension of the Abbott and Deser  construction of a conserved ADM charge 
associated with the Killing vectors of flat spacetime \cite{AD}.  

The construction begins by writing the metric in the form  
$g_{ab} = \eta_{ab} +  h_{ab}$ throughout the spacetime.  The difference
$h_{ab}$ between the full metric and the flat metric may be large in the
interior of the spacetime, but vanishes at the rates specified by  Eqs. (\ref{falloffrates}) 
near transverse spatial infinity. 
Let  $f^{ab}$ be a Yano
tensor for the background metric. Using the Bianchi identities and the 
defining properties of Yano tensors, one can construct a rank $2$ antisymmetric tensor current
$k^{ab}$ which is conserved with respect to the background derivative operator,
{\it i.e.} $ \nabla ^{(0)} _a k^{ab} =0$.  This current is
given by
 \be\label{kab}
k^{ab}= f^{ac}R_c ^{L \  b} -  f^{bc}R_c ^{L \  a}- {1\over 2}f^{ab}R^L
-{1\over 2}   f^{cd}R_{cd} ^{L \  a b} ,
\ee
where $R_{cd} ^{L \  a b}$, for example,  is the linear term in the 
formal power series expansion of $R_{cd} ^{\ \ a b}$
 in $h_{ab}$.   

Conservation of the $2$-form current $k$ can be rewritten as
$d*k=0$, where $*$ indicates the Hodge dual with respect to the background flat 
metric.  It then follows that there locally exists a $3$-form $l$, such that 
$*k=d*l$.  One finds that the $3$-form $l$ is given by
%===========<Equation>===========%
%
\begin{eqnarray}\label{labc}
l^{abc}&=&3\delta^{[a}_k\delta^b_l\delta^c_m \delta^{d]}_nf^{kl}(\bar\nabla^m 
h^n_{\ d})-\frac{1}{4}(\bar\nabla^{[a}f^{bc]})h^n_{\ d}\nonumber\\
       & &+\frac{3}{4}(\bar\nabla^df^{[ab})h^{c]}_{\ d},
\end{eqnarray}  
%
%================================%
As in the previous section, let $V$ be a spacelike,  codimension two slice with normals $n^a$ and $x^a$.
The conserved Y-ADM charge is then defined by the  integral of the $(D-3)$-form 
$*l$ over $\partial V$ as $\partial V\rightarrow\infty$, $i.e.$, over
a closed, codimension two, section of spatial infinity,
%===========<Equation>===========%
%
\begin{eqnarray}
Q[f_{ab}]
=\frac{1}{8\pi}\int_{\partial V}*l,
\end{eqnarray} 
%
%================================%
where we have used the notation $Q[f_{ab}]$ to denote the charge associated with an 
arbitrary asymptotic Killing-Yano tensor $f_{ab}$.

We now focus on the Y-ADM mass density $\mass$, which means specifying the background
Killing-Yano tensor $f^{ab}$ as in equation (\ref{stringyano}) and thereabove.
In this case, since $\bar\nabla_a f^{bc}=0$, only the first term in equation (\ref{labc}) is nonzero.
Substituting the identification (\ref{stringyano}) into this term and carrying out
the anti-symmetrization then yields the expression for the Y-ADM mass density
\be\label{yadcharge} 
\mass = {1\over 8\pi} \int _{\partial V} ds_a q^{ad} q^{cb}\bar \nabla _{[b} \delta g _{d] c}.
\ee
We can further specialize to Cartesian coordinates near infinity, giving the form
\be\label{yadchargecartesian} 
\mass = {1\over 8\pi} \int _{\partial V} ds_I (\partial_J h^{IJ}-\partial^I h_J^{\ J})
\ee
which is very similar to the usual formula for the ADM mass $M$, differing only in the higher codimension of the surface $V$ and the correspondingly smaller range of the indices summed over - here $I,J=2,\dots,D-1$.  

It is interesting to compare the values of the Y-ADM mass density $\mass$ 
and the ADM mass $M_{ADM}$.
As noted above, if the $x^1$ direction tangent to the string is infinite in extent, 
then the ADM mass $M_{ADM}$
will itself be infinite.  The ADM mass will be finite If we consider spacetimes such that $x^1$ compact.  If we make the identification $x^1\equiv x^1 +L$, then the ADM mass is given by
\begin{eqnarray} 
M_{ADM} &=& \int_0^L dx^1 \int_{\partial V} ds_I (\partial_J h^{IJ}-\partial^I h_J^{\ J} \nonumber
-\partial^I h^{\ 1}_1)  \\  & =&
L\, (\mass-\mass_{scalar}),
\end{eqnarray}
where the scalar charge density $\mass_{scalar}$ is defined to by the integral over $\partial_V$ of the 
final term in the integrand above.

It is, lastly, instructive to rewrite the expression for the Y-ADM mass density in terms of $\Lambda_{ab}$, 
the extrinsic curvature of $\partial V$ in $V$. One finds
\begin{equation}
\mass =\frac{1}{16\pi}\int_{\partial V} dv\  (\Lambda^I_{\ I}-\Lambda^{\ \ \ I}_{ ({\rm bg})I  }), 
\end{equation}
where $\Lambda^{IJ}_{({\rm bg})}$ is the extrinsic curvature of $\partial V$ within $V$ 
evaluated with respect to the background flat metric.  
This expression has the same form as that given by
Hawking and Horowitz for the ADM mass in reference \cite{hawking}, 
except that in their work the integral is over the boundary of a codimension one volume.

\section{Spinor Construction of the Y-ADM mass density $\mass$ }\label{spinorsection}

In his proof of the positive energy theorem \cite{Witten}, Witten found an expression for the ADM mass boundary integral in terms of a spinor field.  Stokes' theorem then provides an alternative
volume integral expression for the ADM mass which, providing the spinor satisfies a certain Dirac-type equation on the spacelike hypersurface, is the sum of two terms.  The first of these terms, involving the Einstein tensor, is positive provided the dominant energy condition is satisfied.  The second term, 
quadratic in first derivatives of the spinor field, is manifestly positive and vanishes only in flat spacetime.
Witten's formalism was later streamlined by Nester \cite{Nester}.  

In this section, we develop a generalization of the Witten-Nester formalism that yields a spinor expression for the Y-ADM mass density $\mass$ for a $1$-brane spacetime.  The corresponding
volume integral in this case is again the sum of two terms.  The term quadratic in first derivatives of the spinor field is again positive definite.  However, the analogue of the Einstein tensor term now has contributions from the un-contracted Riemann tensor as well.  Positivity of these terms cannot be ensured by an energy condition.  We explore the issue of conditions that may be imposed on the spacetime to yield positivity of this second term in the volume integral expression for $\mass$.

Our construction begins with finding an appropriate generalization of the Nester $2$-form \cite{Nester}, 
\be\label{oldnester}
B^{ab} = \psi^{\dagger} \gamma^{\hat 0 } \gamma^{abc} \nabla_c \psi  .  
\ee
Within the formalism, 
a $2$-form is the correct object in the ADM case,  because the boundary of a spatial 
hypersurface has codimension $2$ with respect to the full spacetime.   
It is also useful to keep in mind that the ADM mass is defined in terms of the asymptotic Killing vector $\partial/\partial x^0$.  A similar Nester $2$-form was
used in reference \cite{JT2} to study positivity of the ADM tension of a string \cite{JT1,Townsend:2001rg}, which is defined in terms of the asymptotic spatial Killing vector  $\partial/\partial x^1$   tangent to the string.  The Nester $2$-form, in this case, differs  only
 in the substitution of 
$\gamma^{\hat 1}$ for $\gamma^{\hat 0}$ in equation (\ref{oldnester}), 
corresponding to the change in asymptotic Killing vector 
relevant to the two different charges.
 
To compute the Y-ADM mass density $\mass$ for a $1$-brane,
one integrates over the boundary of a co-dimension two spatial surface transverse to the brane.  
This boundary has co-dimension three and we
therefore need to introduce a Nester $3$-form.
Based on the discussion above, we take this to be
\be\label{babc}
B^{abc} = \psi^{\dagger} \gamma^{\hat 0 \hat 1} \gamma^{abcd} \nabla_d \psi .
\ee
We show below that  $\mass$ is indeed given in
terms of this Nester form by
\be\label{boundaryYADM}
 \mass = {1\over 8\pi} \int _{\partial V}ds _a\, n_b\,  y_c\,  ( B^{a bc} + B^{a bc*}) ,
\ee
where $V$ is the co-dimension two surface with normals $n^a$ and $y^a$, as  in 
section (\ref{transversesection}) above, and ${}^*$ denotes complex conjugation.  

We now focus on the volume integral form of $\mass$ obtained from 
equation (\ref{boundaryYADM}) using Stokes theorem \footnote{We assume that there are no horizons
in the spacetime so that no inner boundary terms arise in applying Stokes theorem.  An extension of these results to black brane spacetimes with horizons would presumably require additional boundary 
conditions on the spinor field $\psi$ to eliminate inner boundary terms, as in reference \cite{Gibbons:1982jg}.}
\be\label{stokes}
\mass 
={1\over 8\pi}\int  _V \sqrt{q}\,  n_b\, y_c\, \nabla_a ( B^{a bc} +B^{a bc*}).
\ee
It is then straightforward to show that the volume integrand above may
be  rewritten as the sum of two terms, as in the discsussion of the ADM mass above.
The first term, which involves curvature tensors contracted with bilinears in the 
spinor field $\psi$,  is given by
\bea\label{volone}
(vol)_1 = && R_{ab} n^a \xi ^b -R_{ab} y^a \chi ^b +{1\over 2} R \lambda  \\
&& -R_{abcd}n^a y^b \xi ^c y^d , \nonumber
\eea
where $\lambda= \psi ^\dagger \psi$, 
$\xi ^a = -\psi ^\dagger \gamma^{\hat 0} \gamma ^a \psi$ and  
$\chi ^a ={1\over 2} (\psi ^\dagger \gamma^{\hat 1}\gamma ^a \psi +\, c.c.)$.
The second term, which is quadratic in derivatives of the spinor field,  is given by
\be\label{voltwo}
(vol)_2 = 2\nabla_a \psi^{\dagger} q^{ab}  \nabla_b \psi
  -2(\nabla_a \psi^{\dagger}q^a _{\ c}\gamma ^c ) ( q^b _{\ d}\gamma ^d  \nabla_b \psi )
 \ee
If $\psi$ is a solution to the Dirac-Witten equation $q^a _{\ b} \gamma^b \nabla _
a \psi =0$, then the second term in (\ref{voltwo}) vanishes and $(vol)_2 $ is positive definite. 
The  term $(vol)_1$ may be simplified by choosing the spinor field $\psi$ to be an
eigenvector of $\gammat$with eigenvalue $+i$, which is consistent with the Dirac-Witten equation.
The spinor bilinears in equation (\ref{volone}) are then related according to $\xi^a = -i n^a \lambda$ and 
$\chi ^a = y^a \lambda $.

Combining the two volume terms, we then have the result
\bea\label{spinorone}
\mass &= &{1\over 8\pi }\int _V ( [ R_{ab} n^a n^b -R_{ab} y^a y^b +{1\over 2} R
 \\ && \nonumber
-R_{abcd}n^a y^b n^c y^d  ]\psi ^\dagger \psi  + 2(\nabla_a \psi^{\dagger} )q^{ab}  \nabla_b \psi )
\eea
The volume integrand can be further rewritten in a variety of ways. A compact form which
is useful in subsequent calculations is
\be\label{qqvolone}
(vol)_1 ={1\over 2} R_{abcd} q^{ac}q^{bd} \psidagger\psi
   \ee
 This version highlights the similarities with the codimension one case, see Section VI.
   
A second  form that will also be useful is
 \bea\label{spinortwo}
\mass&  = {1\over 8\pi } \int _V [ &(G_{ab}n^a n^b -R_{abcd}y^a y^c q^{bd})\psidagger\psi \\
&& \nonumber + 2(\nabla_a \psi^{\dagger}) q^{ab}  \nabla_b \psi ]
\eea
Here, after making use of Einstein's equation, 
the first term is simply equal to the energy density $T_{ab}n^a n^b$, which 
supports the idea that the Y-ADM charge 
$\mass$ is a kind of mass per unit length. 
It is less clear how one should interpret 
the Riemann tensor term.  Since it  is not directly related to the stress energy, one cannot
as in the case of the ADM mass and tension, ensure positivity of $\mass$ by imposing an 
energy condition.  However, the fact that the spinor contribution, as in the ADM case, is positive 
definite suggests that one look for  additional conditions on brane spacetimes
such that the Riemann term is also 
positive.  We turn to this issue in Section \ref{positivesection} below.  
First, we return to the issue of the spinor boundary integral in equation (\ref{boundaryYADM}).

\subsection{Further Study of the Spinor Boundary Term}

In this section, we  demonstrate that the spinor boundary integral in equation (\ref{boundaryYADM})
does indeed reproduce the definition of the Y-ADM mass density given in section \ref{YADM}.
We also include a further analysis of the Y-ADM boundary integral, which
shows that the Y-ADM mass density $\mass$ depends only 
on the metric $q_{ab}$ on the surface transverse to the brane. 
This result provides further motivation for regarding the mass density  $\mass $ as 
an intrinsic (rather than extensive) property of the brane.

We consider transverse asymptotically flat spacetimes.  The metric therefore
approaches the Minkowski metric $\eta _{ab}$ at spatial infinity, and 
we can assume that the spinor field $\psi$ approaches one of the constant spinors $\psi^{(0)}$ of
flat spacetime - {\it i.e.} satisfying $\bar\nabla_a  \psizero =0.$ 
The Nester
$3$-form can then be schematically rewritten as $B[\nabla  ]= B[\nabla -\bar\nabla ] + 
B[\bar\nabla] $.  One can check that
$ B[\bar\nabla]$ vanishes sufficiently fast as $\psi$ approaches  $\psi ^{(0)} $ that 
the boundary term just depends only on the difference of the derivative
operators. 
One can then show that 
$(\nabla  _a -\bar\nabla_a  )\psi=-{1\over 4}\nabla _{[m} h_{n] a}\gamma ^{mn}\psi$,  
with the result that  the boundary integrand becomes
\bea\label{btone}
& n_a x_b (B^{abc}  &+ \, c.c. )= \\ &&{1\over 4}  \nabla _{[m} h _{n] d}\, q^{\ c}_a q^{\ d} _b \,
  \psidagger(\gamma ^{ab} 
 \gamma^{mn} +
 \gamma^{mn\dagger} \gamma ^{ab} )  \psi .\nonumber
 \eea
Here we have replaced $\gamma ^{ab\dagger}$  by $\gamma ^{ab}$ since it is 
projected by the spatial metric $q_{ab}$. 
Except for $h_{ab}$, all quantities in equation (\ref{btone}) are background quantities.
Depending on whether the coordinates $x^m , x^n$ are timelike or
spacelike, either the commutator or the anticommutator of the gamma matrices in (\ref{btone})
contributes.
The asymptotically flat boundary conditions (\ref{falloffrates}) imply that
terms involving $h_{xi}$ or $ h_{ti}$, where the indices $i,j$ are
 projected  by $q_{i}^{\ j}$, are higher order.
 It is then straightforward to check that the only terms which contribute to the sum 
in equation (\ref{btone})
 are those with $m,n =i,j$.  We then have 
\be\label{btyano}
\mass ={1\over 8\pi }\int _{\partial V} ds_a q^{ad} q^{cb} \nabla _{[b} h _{d] c}
\ee
which agrees with the expression in section (\ref{YADM}). 

We have seen that the spinor formalism of this section gives an alternate 
construction of Y-ADM charges.  The original formulation \cite{KT} based on the
AD construction \cite{AD} has the merit of establishing a clear relationship between the
Y-ADM charges and the Killing-Yano tensors of the background flat spacetime.  However,
the volume integral form of the charge, in  this construction, involves only the 
linearized curvature tensors around the flat background.  The strength of the 
spinor construction is that it yields a volume integral expression that depends on the exact curvature tensor of the spacetime together with a positive definite spinor term, and is much more useful in
trying to assess positivity properties.

We finish this section by rewriting the Y-ADM mass density $\mass $   in terms of the transverse metric perturbation $\delta q_{ab}$ defined near spatial infinity via the relation 
$q_{ab}=  \delta  _{ab}+\delta q_{ab}$.
Again making use of the assumption that $h_{ti}$ and $h_{xi}$ are of 
higher order at infinity, we find that in the boundary integral (\ref{btyano}) we may write
$  q^b _a q^{cd} q^m  _n \nabla _{b} h _{mc} =D_a \delta q^d _n $, where $D_a$ is the
flat covariant derivative operator on the slice $V$.  The formula for $\mass$ then  becomes
\be\label{btq}
\mass ={1\over 8\pi } \int _{\partial V} ds_a (  D_b ( \delta q_{cd} q^{ac}q^{bd}) - D^a \delta q_{bc}q^{bc} ).
\ee
This expression highlights the fact that $\mass $ is an intrinsic quantity on the codimension-two
volume $V$, since it only depends on $q_{ab}$.
 Also note that if instead $V$ was a codimension-one volume, and $q_{ab}$
the metric on $V$, then equation (\ref{btq}) gives the usual expression for the ADM mass.

\section{Positivity}\label{positivesection}
We have seen above 
that the volume integral expression for the  
Y-ADM mass density $\mass$,  coming from a  rank two Killing-Yano tensor, 
differs from the ADM mass, which comes from a rank one Killing-Yano tensor, in that the Riemann
tensor contributes to the volume integrand, as well as Ricci and scalar curvature 
terms which occur in both cases.  These latter terms can be 
related to the stress energy tensor via the Einstein equation, so that positivity, in the ADM mass 
case,  can be phrased in terms of energy conditions.  The prospects 
for positivity of the Y-ADM mass density $\mass$ therefore look less favorable.
However, the fact that  the spinor field makes a positive definite
contribution to $\mass$,  suggests that we look for conditions under which the Reimann tensor term is
positive as well.  In this section we derive positivity results for the Y-ADM mass density 
$\mass $ under two kinds of conditions; first, 
by requiring conditions on the Weyl tensor of the spacetime,
and second, in the case of exact translational symmetry in the $x^1$-direction.  

\subsection{Conditions on the Weyl Tensor}
The idea here is to decompose the Reimann tensor into trace pieces, the Ricci 
and scalar
contributions, and the rest, the Weyl tensor. When the Weyl term vanishes, the 
remaining terms
in the volume integrand can be related to the stress-energy, which leads to an 
energy condition
for positivity of $\mass $.  In $D$ dimensions, the decomposition of the Riemann tensor is 
given by
\bea\label{weyl}
R_{abcd}& =&C_{abcd} +{2\over (D-2)} ( g_{a[c} R_{d]b} -g_{b[c} R_{d]a} ) \nonumber
\\ &&
+{2\over (D-1)(D-2) } R g_{a[c} g_{d]b}
\eea
Substitution into equation (\ref{volone}) then yields
\bea\label{weylvol}
(vol)_1 &=& [ {D(D-3)\over 2(D-1)(D-2)} R +{D-3\over D-2} 
R_{ab} (n^a n^b -y^a y^b)\nonumber \\ &&
-C_{abcd} n^a y^b n^c y^d  ]\psidagger\psi
\eea
One simple condition to consider is the vanishing of the Weyl tensor, $C_{abcd}=0$.
It is then clear that an appropriate energy condition would assure positivity of $(vol)_1$.

Another condition under which the Weyl tensor term in equation 
(\ref{weylvol}) vanishes is when 
the metric is algebraically special with all four principal null directions coinciding,
and the null vector lies in the plane tangent to the brane.  Explicitly, this means assuming that
$p^a C_{abcd} =0$ for  a null vector  $p^a = n^a \pm y^a$. To see this, define the vectors
$k^a = n^a + y^a$ and
$q^a =n^a -y^a$, which allows us to write
\begin{equation}
C_{abcd} n^a y^b n^c y^d = {1\over 4}C_{abcd} k^a q^b k^c q^d. 
\end{equation}
If either $k^a$ or $ q^a$ is the  principal null vector $p^a$, then this term vanishes. 
Note that while either this conditions or comformal flatness is sufficient to make the
Weyl tensor term in equation (\ref{weylvol}) vanish, but is more than is needed. It would 
be interesting to find a geometrical condition which was minimal in the sense of being both necessary and sufficient.

If the Weyl tensor term vanishes, then $(vol)_1$ can be rewritten in terms of the 
stress energy tensor using Einstein's equation.
If we make the definitions $\rho=T_{ab}n^a n^b$, $p_{\hat y} =T_{ab} y^a y^b$ and similar 
definitions for the
pressures $p_{\hat i}$ in directions transverse to the string, then the volume integral
expression for $\mass$ becomes
\bea\label{zeroweyl}
\mass & =&   {(D-3) \over  (D-1)}\int _V [  (\rho - p_{\hat y} 
 +{1\over D-2}\sum_i  p_{\hat i})\psidagger\psi \nonumber
 \\ &&
 +2 (\nabla _a \psidagger )q^{ab}\nabla _b \psi ]
\eea
 where the index $i$ runs over the transverse coordinates. 
 If we therefore have any set of condition, such as those suggested above, 
 which lead to the vanishing of the Weyl tensor term in (\ref{weylvol}), then
 we have shown that the Y-ADM mass density $\mass$ is positive if the additional energy condition
\be\label{energycond}
(\rho - p_{\hat y}  +{1\over D-2}\sum_i p_{\hat i}) \geq 0
\ee
 is satisfied \footnote{This energy conditions differs from standard energy conditions, such as the 
dominant energy condtion, in that it refers to the normal vectors in a given slicing of the spacetime, 
rather than a general class of vectors.}.
 
The energy condition (\ref{energycond}) has a straightforward physical interpretation.
The quantity $-p_{\hat y}$ is simply the tension along the string, and equation (\ref{zeroweyl})  implies 
that this tension contributes positively to $\mass$.  Note that for a standard cosmic string we have
$p_{\hat y}= -\rho $.
The pressures $p_{\hat i}$ in the
directions transverse to the string
contribute as they do for a particle-type configuration -  positive pressures make a positive
contribution to the mass.   Note also that in the non-relativistic limit, where the pressures are negligible compared to the energy density, equation (\ref{zeroweyl}) 
implies that  $\mass$ is the energy density integrated over a transverse slice, plus
a positive definite contribution from the gravitational field.

%\subsection{Relation of the Y-ADM Mass Density $\mass$ to the ADM Mass $M$}
%Another approach to addressing positivity properties of the Y-ADM mass density $\mass$ is to look
%at its relation to the ADM mass $M$.  We can then appeal to the known positivity properties of $M$.  AS noted above, if the $x^1$ direction along the string is infinite in extent, then the ADM mass itself is infinite.  Hence, in this section we take the $x^1$ direction to be compact with the identification $x^1\equiv x^1+L$.  There are then two comparisons we can make.  We can look at the relation of the 
%Y-ADM mass density $\mass$ to the $D$-dimensional ADM mass $M^{(D)}$.  We can also look at the relation between $\mass$ and the $(D-1)$-dimensional ADM mass $M^{(D-1)}$ defined with respect to a Kaluza-Klein reduction.

%The $D$-dimensional ADM mass is given by the formula
%%
%It is reasonable to ask what is the relation between the mass density $\mass$ and the 
%ADM mass. There are two distinct cases-- the $(D-1)$-dimensional ADM mass $M^(D-1)$, and
%the $D$-dimensional ADM mass $M^D$. The first case is simpler, so let
%$g_{ab}= -n_a n_b +q_{ab}$ be a $(D-1)$-dimensional
%asymptotically flat spacetime.   The mass is an integral over a $(D-3)$-dimensional sphere
%at infinity,
%%
%\be\label{lowermass}
%M^(D-1)  ={1\over 8\pi } \int _{\infty} ds_a (  D_b ( \delta q_{cd} q^{ac}q^{bd}) - D^a \delta q_{bc}q^{bc} )
%\ee
%which is the same as (\ref{btq}).

\subsection{Translational Symmetry and Kaluza-Klein Reduction}
Suppose that translation along the string is a actually a symmetry of the $D$-dimensional spacetime. One would then expect that there is a $(D-1)$-dimensional point of view, in which the Y-ADM mass
density $\mass$ is related to the $(D-1)$-dimensional ADM mass. 
%For example, if we take the direction along the string to be compact and carry out 
%a Kaluza-Klein reduction, then the $D$-dimensional theory is
%reduced to  $(D-1)$-dimensional Einstein gravity coupled to a scalar dilaton and a $U(1)$ gauge field, for which the ADM mass is positive.  However, even with the rather strong assumption that there is a Killing field, the situation is richer than just  the Kaluza-Klein case suggests. 
%We will give positivity results in two kinds of situations;  first, when a Kaluza-Klein reduction is appropriate, and second, when the point of view is still intrinsically $D$-dimensional.

We will need the following Gauss-Codazzi relation for a codimension one submanifold with unit normal vector $w^a$.  If we write the metric as
\begin{equation}
g_{ab}=s_{ab} +(w\cdot w) w_a w_b,
\end{equation}
where $ w\cdot w =\pm 1$
depending on whether the normal  $w^a$
spacelike or timelike, then the Riemann tensors of $g_{ab}$ and $s_{ab}$ are related according to
\bea\label{gausscod} 
s^m_a s^n_b s^r_c s^p_d R_{mnrp}[g] & = &R_{abcd}[s] \\ &&\nonumber
+(-w\cdot w)( J_{ac} J_{bd} - J_{ad} J_{bc}) 
\eea
where $J_{ab}=s_a ^{\ c} \nabla _c w_b$ is the extrinsic curvature of the submanifold. 

%We first consider the case of  Kaluza-Klein reduction, with 
Take the normal vector to be $y^a$, the 
direction tangent to the string.  The metric $s_{ab}$ is then Lorentzian.  The volume integrand for $\mass$ can be expressed in terms of the Einstein tensor for a $(D-1)$-dimensional metric  
$G_{ab}^{(D-1)} [s] $ as follows.  Making use of equation (\ref{gausscod}) one finds that
 \bea\label{bigequation}
 (vol)_1 &=&{1\over 2} q^{ac} q^{bd} R_{abcd} [g]\nonumber\\
 &=& {1\over 2} q^{ac} q^{bd}( R_{abcd}[s]
 +  J_{ac} J_{bd} - J_{ad} J_{bc}) \\ &=&
 G_{ab}^{(D-1)}  [s ]  n^a n^b + {1\over 2} (\beta _{ab} \beta ^{ab} -\beta ^2 )\nonumber
\eea
where
$\beta _{ab}= q_a ^{\ m} q_b ^{\ n} \nabla _m y_n$ is the projection of the extrinsic curvature 
$J_{ab}$ onto the directions transverse to the string.

We now assume that the spacetime has a spatial translation Killing field $V^a$ that  is parallel
to the brane.  Near infinity, 
the Killing vector $V^a$ approaches $\partial /  \partial x^1$, which also coincides with 
the unit normal vector $y^a$ in this limit.
Further assume that we extend $y^a$ into the interior of the spacetime, 
such that it is parallel to the Killing field $V^a$,  {\it i.e.}  let
 $V^a =F y^a$ where $V_a V^a =F^2$.
We then have $J_{ab}=0$, since $J_{ab}  =(1/2) \mbox \pounds_V s_{ab}$.
From the last line of equation (\ref{bigequation}), we then have the result
\be\label{kkmass}
\mass   = {1\over 8\pi }\int  _V \left (n^a n^b G^{(D-1)} _{ab}[s ]\psidagger \psi +
 2(\nabla_a \psi^{\dagger} )q^{ab}  \nabla_b \psi \right)
\ee
At this point one needs to know something about the dimensionally 
reduced theory \footnote{Note that the metric $s_{ab}$ is not rescaled by a dilaton factor, as it would be in the Kaluza-Klein ansatz, and that no Kaluza-Klein gauge fields appear because we have chose the normal direction to be
parallel to the Killing vector $v^a$.  Therefore, in these coordinates, the shift terms $g_{yt}$ and $g_{yI}$ vanish.}, in order to draw a conclusion 
about the Y-ADM mass density $\mass$.  
If the dimensionally reduced Einstein tensor is equal to a stress energy
 which satisfies the dominant energy condition, then $\mass$ is positive. Indeed, the form of the boundary term given in equation (\ref{btq}) is the standard ADM mass for the metric
$s_{ab}$.  Therefore, the argument leading to (\ref{kkmass}) gives a consistency check on the
meaning of the mass per unit length: when dimensional reduction is possible, the mass per
unit length coincides with the ADM mass of the lower dimensional theory.

\section{A Generalized Hamiltonian for Transverse Asymptotically Flat Spacetimes?}
In this final section we point out that the form of the volume integral expression for $\mass$ derived in section \ref{spinorsection} suggests that the term $(vol)_1$ might serve as a generalized Hamiltonian for the evolution for transverse asymptotically flat brane spacetimes.  By generalized Hamiltonian evolution we mean specifying data on a codimension $2$ slice transverse to the brane and evolving the data via a system of first order PDE's in the directions tangent to the brane.
In this view, we think of ordinary Hamiltonian evolution as specifying data on a codimension $1$ slice transverse to the worldline of particle-like sources and then evolving in the single direction tangent to
the particle worldline.

First recall that in the codimension $1$ case, if we write $g_{ab} =-n_a n_b +l_{ab}$, 
then the Einstein tensor term in the volume integrand is simply given by 
\begin{eqnarray}
(vol)_1 &=& (G_{ab} n^a n^b) \psidagger\psi\\
 &=&{1\over 2} (R_{abcd} l^{ac} l^{bd})\psidagger\psi \label{hamform}
\end{eqnarray}
If we then change to  Hamiltonian variables we have
\begin{equation}
(vol)_1={1\over 2} (  R[ l] - k_{ab}k ^{ab} + k^2 )\psidagger \psi
\end{equation}
where $R[ l]$ is the scalar curvature of the $D-1$ dimensional metric $l_{ab}$
and $k_{ab}= l_a ^{\ c} l_b ^{\ d} \nabla _c n_d$ is the extrinsic curvature of the 
slice with normal $n_a$.  The factor multiplying $\psidagger \psi$ is, of course, simply proportional
to the gravitational Hamiltonian.  An additional momentum term would also arise if $\psi$ were not taken to be an eigenvector of $\gamma^{\hat 0}$.

In the  codimension $2$ case addressed in this paper, it is interesting  
that the volume term $(vol)_1$ has a very similar form to equation (\ref{hamform}) and may be similarly
rewritten in Hamiltonian form, 
\bea\label{qqvol}
(vol)_1 &=& {1\over 2}( R_{abcd} q^{ac}q^{bd} )\psidagger\psi \\
 &=& {1\over 2}( R[q] - K_a ^{ae} K^a _{\  ae} + K_{ab}^{\ \  e} K^{ab}_{\ \  e})
 \psidagger\psi\nonumber
\eea
It seems natural to speculate that this quantity, along with the analogous momentum terms that would
appear if we did not take $\psi$ to be an eigenvector of $\gamma^{\hat 0}$, play a role in a generalized
Hamiltonian evolution of brane spacetimes of the sort described above.
We note, as well, that extrinsic curvature can be further decomposed as $K_{ab}^c =\alpha _{ab} n^c +\beta _{ab}y^c$,
where $\alpha _{ab} = q_a^c q_b^d \nabla _c n_d$ and  
$\beta _{ab} = - q_a^c q_b^d \nabla _c y_d$.
We can then write
\bea\label{hamiltonianl}
{1\over 2} R_{abcd} q^{ac}q^{bd} &=&  {1\over 2} ( R[ q] - \alpha_{ab}\alpha ^{ab} + \alpha ^2  \\
&&\nonumber + \beta_{ab}\beta ^{ab}  
-   \beta ^2 ),
\eea
where all  contractions are done with the positive 
definite metric $q_{ab}$.  We can thus one can read off whether 
the extrinsic curvature terms are positive or negative.

It is also interesting to compare equation (\ref{qqvol}) to the Hamiltonian in the double null formalism developed by Hayward \cite{Sean:1993}.  
In certain gauge conditions, equation (\ref{qqvol}) is identical with 
the Hamiltonian in double null formalism, at least in four dimensional spacetimes (see reference \cite{Sean:1994}).  The extension to higher dimensional spacetimes is straightforward.
Since we focus on the codimension $2$ integral manifolds, the twist term also 
vanishes.  Therefore, we can realise that  equation (\ref{qqvol}) is indeed the Hamiltonian. 
This is a natural consequence because the double null formalism provides us with a codimension $2$ space normal to two null directions.

\vskip 0.3cm
\noindent{\bf Acknowledgements:} 
The work of DK and JT was supported in part by National Science Foundation grant NSF PHY0244801. 
The work of TS was supported by Grant-in-Aid for Scientific 
Research from Ministry of Education, Science, Sports and Culture of 
Japan(No.13135208, No.14740155 and No.14102004). 
The work of ST was supported 
by the 21st Century COE Program at TokyoTech "Nanometer-Scale Quantum Physics" supported 
by the Ministry of Education, Culture, Sports, Science and Technology.


\begin{thebibliography}{22}


\bibitem{KT}
D.~Kastor and J.~Traschen,
``Conserved gravitational charges from Yano tensors,''
JHEP {\bf 0408}, 045 (2004)
[arXiv:hep-th/0406052].
%%CITATION = HEP-TH 0406052;%%

\bibitem{Yano}
K.~Yano, ``Some Remarks on Tensor Fields and Curvature," Ann.\ Math.\ {\bf 55}, 328 (1952).

\bibitem{AD}
%\cite{AD}
L.~F.~Abbott and S.~Deser,
``Stability Of Gravity With A Cosmological Constant,''
Nucl.\ Phys.\ B {\bf 195}, 76 (1982).
%%CITATION = NUPHA,B195,76;%%

%\cite{Schon:1979rg}
\bibitem{Schon:1979rg}
R.~Schoen and S.~T.~Yau,
``On The Proof Of The Positive Mass Conjecture In General Relativity,''
Commun.\ Math.\ Phys.\  {\bf 65}, 45 (1979).
%%CITATION = CMPHA,65,45;%%

\bibitem{Witten}
E.~Witten,
``A Simple Proof Of the Positive Energy Theorem,''
Commun.\ Math.\ Phys.\  {\bf 80}, 381 (1981).
%%CITATION = CMPHA,80,381;%%

\bibitem{Nester}
J. Nester, Phys. Lett. {\bf 83A}, 241(1981). 

%\cite{Townsend:2001rg}
\bibitem{Townsend:2001rg}
P.~K.~Townsend and M.~Zamaklar,
``The first law of black brane mechanics,''
Class.\ Quant.\ Grav.\  {\bf 18}, 5269 (2001)
[arXiv:hep-th/0107228].
%%CITATION = HEP-TH 0107228;%%

\bibitem{JT2}
J.~H.~Traschen,
``A positivity theorem for gravitational tension in brane spacetimes,''
Class.\ Quant.\ Grav.\  {\bf 21}, 1343 (2004)
[arXiv:hep-th/0308173].
%%CITATION = HEP-TH 0308173;%%

\bibitem{JT1}
J.~H.~Traschen and D.~Fox,
``Tension perturbations of black brane spacetimes,''
Class.\ Quant.\ Grav.\  {\bf 21}, 289 (2004)
[arXiv:gr-qc/0103106].
%%CITATION = GR-QC 0103106;%%

%\bibitem{SIT}
%T. Shiromizu, D. Ida and S. Tomizawa, Phys. Rev. {\bf D69}, 027503(2004). 

%\bibitem{HO}
%T. Harmark and N. A. Obers, JHEP {\bf 0405}, 043(2004).



\bibitem{hawking}
S.~W.~Hawking and G.~T.~Horowitz,
``The Gravitational Hamiltonian, action, entropy and surface terms,''
Class.\ Quant.\ Grav.\  {\bf 13}, 1487 (1996)
[arXiv:gr-qc/9501014].
%%CITATION = GR-QC 9501014;%%




%\bibitem{SMS}
%T. Shiromizu, K. Maeda and M. Sasaki, Phys. Rev. {\bf D62}, 024012(2000).

%\bibitem{GH}
%G.~W.~Gibbons and C.~M.~Hull,
%``A Bogomolny Bound For General Relativity And Solitons In N=2 Supergravity,''
%Phys.\ Lett.\ B {\bf 109}, 190 (1982).
%%%CITATION = PHLTA,B109,190;%%


%\cite{Gibbons:1982jg}
\bibitem{Gibbons:1982jg}
G.~W.~Gibbons, S.~W.~Hawking, G.~T.~Horowitz and M.~J.~Perry,
``Positive Mass Theorems For Black Holes,''
Commun.\ Math.\ Phys.\  {\bf 88}, 295 (1983).
%%CITATION = CMPHA,88,295;%%


\bibitem{Sean:1993}
S. A. Hayward, Class. Quantum Grav. {\bf 10}, 779(1993).

\bibitem{Sean:1994}
S.~A.~Hayward,
``Quasilocal gravitational energy,''
Phys.\ Rev.\ D {\bf 49}, 831 (1994)
[arXiv:gr-qc/9303030].
%%CITATION = GR-QC 9303030;%%

\end{thebibliography}
\end{document}